\documentstyle[graphics,psfig,epsf,amssymb]{aa}
\begin{document}
\thesaurus{02(02.08.1, 02.09.1, 02.20.1)}
\title{Turbulence in differentially  rotating  flows}
             \subtitle{What can be learned from the Couette-Taylor experiment}
            \author{Denis Richard and Jean-Paul Zahn}
             \offprints{D.Richard, Denis.Richard@obspm.fr}
             \institute{D\'epartement d'Astrophysique Stellaire et Galactique/CNRS-UMR8633,  \\
Observatoire de Paris-Meudon, 92195 Meudon Cedex, France \\}

\date{Received : November 25 1998 ; accepted : March 24 1999}	

\maketitle

\begin{abstract}

The turbulent transport of angular momentum plays an important role in many astrophysical
objects, but its modelization is still far from satisfactory.
We discuss here what can be learned from laboratory experiments.
We analyze the results obtained by Wendt (1933) and Taylor (1936) on the classical
Couette-Taylor flow, in the case where angular momentum increases with
distance from the rotation axis, which is the most interesting for astrophysical
applications. We show that when the gap between the coaxial cylinders is wide enough, 
the criterion for
the onset of the finite amplitude instability can be expressed in terms of a gradient 
Reynolds number.
Based on Wendt's results, we argue that turbulence may be sustained by differential rotation
when the angular velocity decreases outward, as in keplerian flows.
From the rotation profiles and the torque measurements we deduce a
prescription for the turbulent viscosity which is independent of gap width;
with some caution it may be applied to stellar interiors and to accretion disks.


\end{abstract}

\section{Introduction}

Differential rotation is observed in various astrophysical objects, 
from planets to galaxies, and one suspects that it gives rise to turbulence, since shear
flows are liable to hydrodynamical and MHD instabilities. When these instabilities are
of the linear type, they are relatively easy to study by perturbing slightly the
equilibrium state. But some of them occur only at finite amplitude, in which
case the answer must be sought in computer simulations or laboratory experiments, with
their inherent limitations. 

There has been some debate recently on whether a keplerian disk, which is linearily
stable (Rayleigh 1916), may be unstable to finite amplitude
perturbations.  It may look as if this question presents little interest, since it has
been proved that a very weak magnetic field suffices to
render such a disk linearly unstable (Chandrasekhar 1960; Balbus and Hawley 1991).
 However the properties of angular
momentum transport depend sensitively on which instability dominates in the considered
regime, and a finite amplitude instability can overpower the linear instability which is
the first to occur, as the relevant control parameter increases. One example is the
Couette-Taylor flow, with the outer cylinder at rest and the inner cylinder rotating
with angular velocity $\Omega$. 
  When $\Omega$ is increased, the transport of angular
momentum first scales as $\Omega^{3/2}$,
but thereafter it varies as $\Omega^2$, once the flow has
become fully turbulent (Taylor 1936),  as if the initial linear instability were
superseded by a stronger shear instability (see also Lathrop et al. 1992). 
By extrapolating the $\Omega^{3/2}$ law to high $\Omega$
one would clearly underestimate the transport. 

In the present article, we take as working assumption that any	
differentially rotating flow experiences, at high Reynolds number, the turbulent regime
observed in the Couette-Taylor (CT) experiment, and that this turbulence will then 
transport angular momentum in the same way as in that experiment. 
The CT flow has been chosen as		
reference because it is the simplest flow to realize in the laboratory, with both shear
and rotation that can be varied independently. We examine whether the experimental
data suggest a prescription for the angular momentum transport, which may be used to
model astrophysical objects. A similar approach has been taken by Zeldovich (1981), but
our conclusions will differ from his (see Appendix).

\section{The Couette-Taylor experiment}

The CT experimental apparatus consists of two coaxial rotating 
cylinders of radius $R_1$ and $R_2$ separated by a gap $\Delta R = R_2 - R_1$, which is 
filled with a fluid of viscosity $\nu$. The
 cylinders can rotate with different angular velocities $\Omega_1$ and $\Omega_2$;
their height is in general much larger than their radius, to minimize
the effect of the boundaries. The Reynolds number is usually  defined in terms of the 
differential rotation $\Delta \Omega = |\Omega_2 - \Omega_1|$ and  by taking
the gap width as characteristic length:
\begin{equation}
Re = {\Delta \Omega \, R \, \Delta R \over \nu} ,
\end{equation}
where $R$ is the mean radius $ R = (R_1 + R_2)/2$.

\begin{figure}
\begin{center}
\leavevmode
\psfig{figure=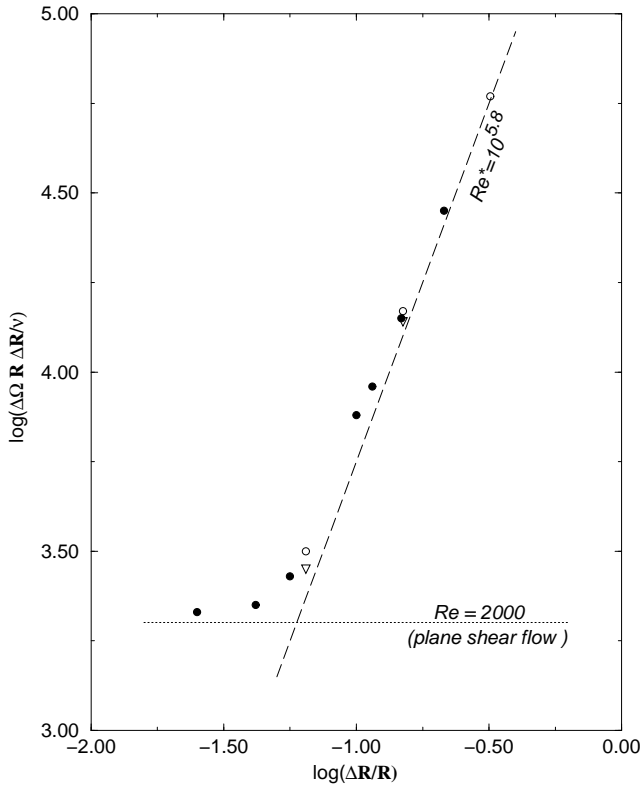,width=8.5cm}
\end{center}
\caption{ Critical Reynolds number vs. aspect ratio; filled circles from Taylor (1936),
open circles from Wendt (1933). Dotted line: critical Reynolds number for plane shear flow
instability; dashed line: critical gradient Reynolds $Re^*$ for circular shear flow in the limit
of large gap (see text). 
 }
\end{figure}

\subsection{Critical parameters and transition to turbulence}

When the inner cylinder is rotating and the outer one is at rest, angular momentum 
decreases outward and the flow is linearly unstable for Reynolds numbers
higher than $Re_c = 41.2 \, \sqrt{R/\Delta R}$ (for narrow gap, cf. Taylor 1923). This well
known instability takes first the shape of steady toroidal, axisymmetric cells (the
Taylor vortices); it is very efficient in transporting angular momentum, whose gradient
is strongly  reduced. At higher $Re$ a wavy pattern appears and after a series of
bifurcations the flow becomes fully turbulent. This case has been
studied by many experimental teams, and it is extremely well documented (see Andereck
et al. 1986). It has also been modeled successfully in three-dimensional numerical
simulations (Marcus 1984; Coughlin \&  Marcus 1996).

In the opposite case, when the outer cylinder is
rotating and the inner one is at rest, the angular momentum increases outward and the
flow is linearly stable. The only theoretical prediction concerning the
non-linear behavior is that by Serrin (1959), later refined by Joseph and Munson (1970),
who established that the flow is stable
against finite amplitude perturbations below $Re = 2 \pi^2$ (in the narrow gap limit).


To this date, no numerical
simulation has been able to demonstrate the finite amplitude instability.
But this instability does occur in the laboratory, and it has been described already by
Couette (1890). It was studied in detail by Wendt (1933) and Taylor (1936), who showed that for
Reynolds numbers exceeding $Re_c \simeq 2\, 10^3$, the flow becomes unstable and
immediately displays turbulent motions. The critical Reynolds number depends on
whether the angular speed is increased or decreased in the experiment, a typical property
of finite amplitude instabilites. Moreover, it is sensitive to gap width, as demonstrated
by Taylor. 
Figure 1 displays results from Wendt and Taylor: 
$Re_c$ is roughly constant below $\Delta R / R = 1/20$, but above it increases as
$\approx (\Delta R / R)^2$, as was already noticed by Zeldovich (1981),
a behavior for which an explanation was proposed by Dubrulle (1993). In the latter regime
one can define another critical
Reynolds number $Re_c^*$ involving, instead of gap width, the gradient of angular
velocity; since  
\begin{equation} Re_c = {R^3 \over \nu} {\Delta \Omega \over \Delta R}  
\left({\Delta R \over R }\right)^2 = Re^*_c \left({\Delta R \over R}\right)^2 ,
\end{equation}
the instability condition becomes
\begin{equation}
Re^* = {R^3 \over \nu} {\Delta \Omega \over \Delta R} \geq Re^*_c \simeq 6 \, 10^5.
\end{equation}
We see that two conditions must be satisfied for the finite amplitude instability to
occur: the first $Re \geq Re_c$ is the classical criterion of shear instability, valid
also for plane parallel flows, whereas the second $Re^* \geq Re^*_c$, involving 
what we shall call the
{\it gradient} Reynolds number, is genuine to differential rotation.
In addition, to trigger the instability the strength of the perturbation must exceed a 
certain threshold, which presumably also depends on $Re$ or $Re^*$.

\subsection{Transport of angular momentum}

In the turbulent regime, the torque measured by Taylor scales approximately as
$G \propto (\Omega_2)^n$ for a given gap width, where the exponent $n$ tends to 2 for
large $\Omega_2$.   The measurements made by Wendt confirm that scaling with $n \approx 2$.
 It suggests that the transport of angular momentum may be considered as a
diffusive process, and that the mean turbulent viscosity $\overline \nu_t$ increases 
linearly with
$\Omega_2$, or $\Delta \Omega$. It is then natural to examine whether this
viscosity may be expressed as

\begin{figure}
\begin{center}
\leavevmode
\psfig{figure=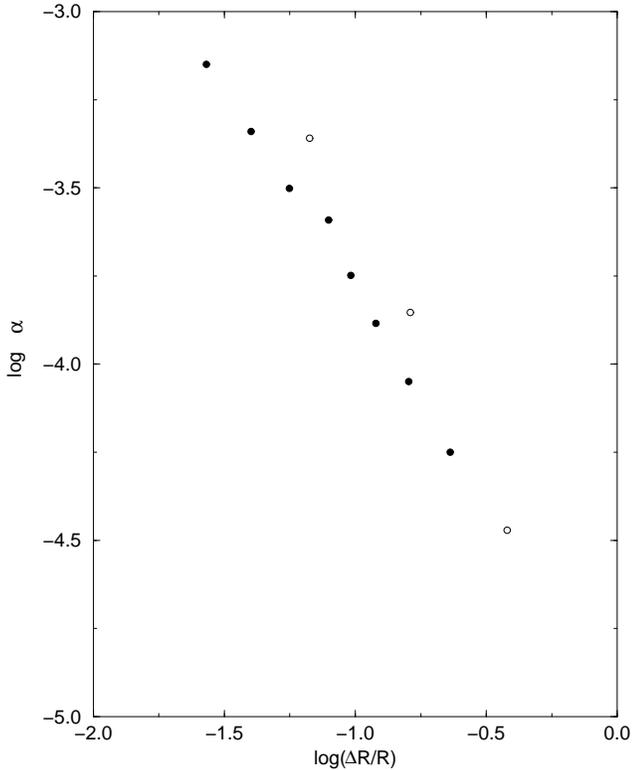,width=8.5cm}
\end{center}
\caption{ Value of parameter $\alpha$ in the classical viscosity prescription 
$\nu_t = \alpha R \,
 \Delta \Omega \, \Delta R$, derived from  Taylor (1936) in filled circles and from
 Wendt (1933) in open circles.}
\end{figure}

\begin{equation}
\overline \nu_t = \alpha R \, \Delta \Omega \, \Delta R
\label{alpha}
\end{equation}
with $\alpha$ being a constant of order unity, since the largest turbulent eddies
would have a size $\approx \Delta R$ and a peripheral velocity $\approx R \, \Delta
\Omega /2$. The parameter $\alpha$ is easily derived from the torque measurements, and
the surprising result is that it decreases with gap width (fig. 2). For the smallest gaps,
$\alpha$ scales as the inverse of $\Delta R/R$, but the slope steepens farther 
as if the scaling would tend asymptotically to 
\begin{equation}
\alpha \propto \left(\Delta R \over R \right)^{-2} \qquad \hbox{for} \,  
\left(\Delta R \over R \right) \rightarrow 1 .
\end{equation}
(Comparing Wendt's experimental data of his two largest gaps, one deduces an exponent
$\delta \ln \alpha / \delta \ln (\Delta R/R_2) = - 1.83.$) 

We may therefore conclude that, in the limit of large gap, the mean turbulent viscosity 
actually scales as 
\begin{equation}
\overline \nu_t \simeq \left( {\alpha \, \Delta R \over R} \right)^2 R^3  
{\Delta \Omega \over \Delta R}
= \beta^* R^3 {\Delta \Omega \over \Delta R} ,  
\end{equation}   
with $\beta^* \approx 4 \, 10^{-6}$.

This strongly suggests that the {\it local} value of the turbulent viscosity is then
independent of gap width, and that it is determined only by the local shear:
\begin{equation}
\nu_t = \beta r^3 \left|{d  \Omega \over d  r}\right| ,
\label{local}
\end{equation}
$r$ being the radial coordinate.

In principle, one should be able to verify this prescription for $\nu_t$ by examining the 
rotation
profiles measured by Taylor and Wendt. According to (\ref{local}), the 
conservation of angular momentum requires that its flux,
given by
\begin{equation}
{\cal F} = \left[ \nu + \left|\beta r^3 {d  \Omega \over d  r} \right|\right] \, r^2   
{d  \Omega \over d  r} ,
\label{flux}
\end{equation}
varies as $1 / r$ between the cylinders. Therefore $r^3 (d  \Omega / d  r)$ 
should be constant in the turbulent part of the profile (as it is in the laminar flow).
 
But this constancy can be expected only if the transport of angular momentum is achieved
by the viscous and turbulent stresses alone. That is not the case in Taylor's experiment: 
as acknowledged by him, an Ekman circulation is induced by the ends of his apparatus, 
although 
he tries to minimize the boundary effects by chosing a large aspect ratio (height/radius). 
Moreover his rotation profiles are deduced from pressure measurements made with a Pitot tube 
located at half
height of the cylinders, where the radial return flow has its maximum intensity. Consequently,
the torque inferred 
from these profiles is about half of that measured directly at the inner cylinder.

The aspect ratio is less favorable in Wendt's experiment, but there the top boundary is
a free surface, and most of his measurements have been 
made with the bottom boundary split in two annuli, attached respectively to the inner and
the outer cylinders, which reduces drastically the circulation and
renders his results more reliable. We examined his rotation profiles obtained 
for the largest gap width ($\Delta R / R = 0.38$) and
with 4 different speeds of the inner cylinder ($0 \leq \Omega_1 \leq \Omega_2/2$);
we found that in the bulk of the fluid, 
these profiles are compatible with the constancy
of the gradient Reynolds number, as predicted by (\ref{flux}).
However we cannot rule out a mild variation of $\beta$ within the profile.
Unfortunately, Wendt gives the torque only for the case where the inner 
cylinder is at rest;
we draw from it the following value of the coefficient $\beta$:
\begin{equation}
\beta = 1.5 \pm 0.5 \, 10^{-5} .
\label{beta}
\end{equation}
The estimated uncertainty reflects that of the velocity measurements:
$\beta$ results from second derivation of the shape of the fluid surface.

\section{Are keplerian flows unstable?}


\begin{figure}
\leavevmode
\rotatebox{270}{
\psfig{figure=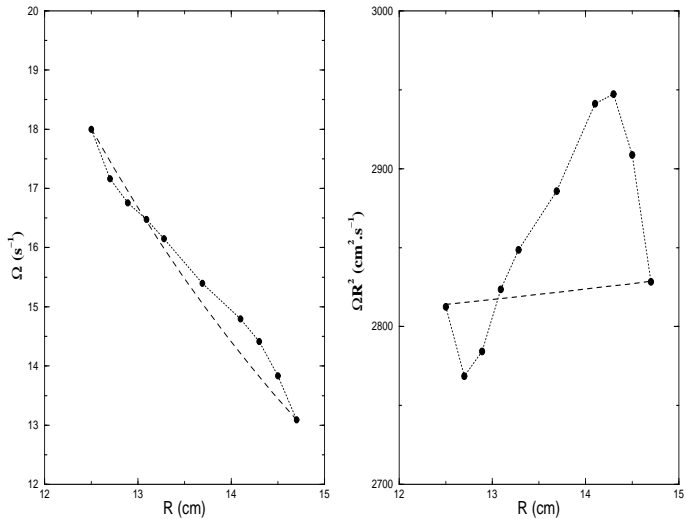,width=7cm,height=9cm}}
\caption{ Angular velocity and momentum profiles in the case of decreasing
angular velocity and initially constant angular momentum -- dotted line (experimental data
from Wendt 1933). }
 \end{figure}

More precisely, the question we address is whether  the Couette flow is unstable when the 
angular velocity decreases outwards and the 
angular momentum increases outwards, as in keplerian flow.

There is no definite answer yet, because this regime has not been explored at high enough Reynolds 
number. But some information can be gleaned from Wendt's study. He reports the results of
experiments where the two cylinders rotate such that
$\Omega_2 R_2^2 = \Omega_1 R_1^2$. At low Reynolds number, this setup enforces a laminar
flow of constant angular momentum (neutral flow), but at high Reynolds number this flow becomes
turbulent for two of the three gaps used by Wendt. 

The angular velocity and angular momentum profiles for one of these turbulent flows
are reproduced in fig.3. 
(According to Wendt's data,
$ \Omega_2  R_{2}^2$ actually exceeds $\Omega_1  R_{1}^2$ by a half a percent.) 
The profiles clearly demonstrate the flow instability, with angular momentum being transported
down the angular velocity gradient, which becomes somewhat flater far enough
from the boundaries,  whereas the angular momentum profile steepens substantially.

In the turbulent bulk of the flow $q = - {d\ln \Omega / d\ln r } \approx  1.4$, compared
to the initial $q = 2$. We recall
that $q = 1.5$ in keplerian flow, and 
that in the numerical simulations performed by Balbus et al. (1996, 1998), the instability is lost
already at about $q = 1.95$.  A crude estimate of the parameter $\beta$ in the viscosity formulation
 (\ref{local}) indicates
that the size of the turbulent eddies is much smaller than the gap width: 
$\ell = \sqrt{\beta} R \approx \Delta R / 100$.

The corresponding values of $Re^*$ for the three experiments are reported  in fig.4 together 
with the critical line of fig.1.
They are located respectively above and below this line for the unstable and stable flows. The 
data are too scarce to locate precisely the critical line of these ``neutral'' flows, but we can 
conclude that the critical gradient Reynolds number $Re^*$ then lies between $2 \, 10^5$ and $6 \, 10^5$.

\begin{figure}
\begin{center}
\leavevmode
\psfig{figure=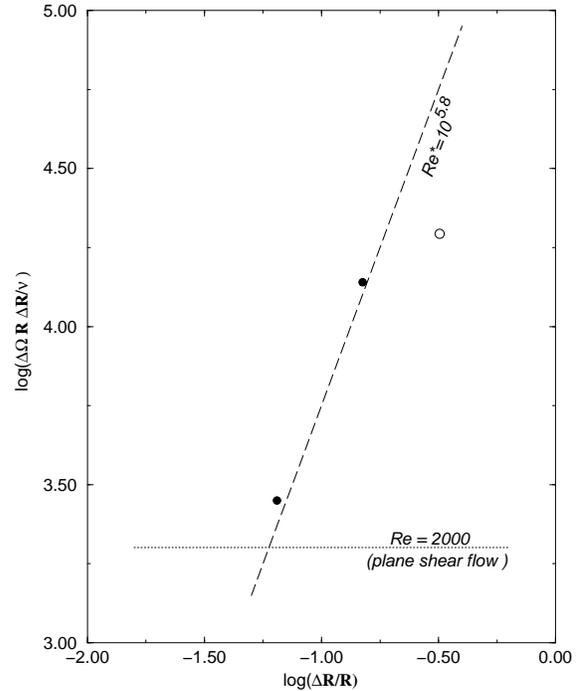,width=7.5cm}
\end{center}
\caption{  Reynolds number vs. aspect ratio for the three ``neutral'' experiments from Wendt: 
the two filled circles correspond to the unstable flows, the open circle to the stable flow.
The stability curve obtained for the inner cylinder at rest (fig.1) is displayed for comparison.}
\end{figure}

\section{Discussion}

The firmest result of our analysis of the Couette-Taylor experiment is that 
the criterion for finite amplitude instability 
may be expressed in terms of the gradient Reynolds number
$Re^* = R^3 (\Delta \Omega / \Delta R)/\nu$, where the critical 
Reynolds number $Re^*_c \lesssim 6 \, 10^5$
is independent of the width of the gap between cylinders, for wide enough gap.
The turbulent transport of angular momentum then seems also to be independent of gap width;
it proceeds always down the angular velocity gradient, as confirmed by the behavior of
the initially ``neutral'' flows examined by Wendt.

Though the experimental evidence is somewhat less compelling,
we have established empirically an expression which links the turbulent viscosity
to the local shear.
The value of $Re^*_c$, and that of $\beta$ in eq.(\ref{local}), have been derived from Wendt's
experiment with the inner cylinder at rest, and it is not obvious that these parameters would 
be the same for different ratios of cylinder speeds. Also, the linear scaling $\nu_t/\nu = \beta
Re^*$ may be valid only for those moderate gradient Reynolds numbers which could be reached
in the laboratory

Nevertheless, it is tempting to apply this expression (\ref{local})
to accretion disks, as an alternate 
for the commonly
used prescription $\nu_t = \alpha c_s H$, where $H$ is the scale height of the disk and 
$c_s$ the local sound speed (Shakura \& Sunyaev 1973). 
Some caution is required because this viscosity has been derived from experiments 
performed with an incompressible fluid.
Moreover (\ref{local}) implies that the eddies which dominate in
the transport of angular momentum have a size of order $\ell \approx \sqrt\beta \, r$, 
independent of 
the strength of the local shear, and that their velocity is of order $\ell  r \, d\Omega/dr$.
When applying this prescription to a compressible flow, one has to make sure that this
velocity is smaller than the sound speed and, in the case of an accretion disk,
that the size $\ell$ of the eddies, which are three-dimensional, does not largely exceed 
the scale height $H$. The behavior of ``neutral'' flows demonstrates that the shear instability
always transports angular momentum down the angular velocity gradient, which means outward for accretion disks.


Note that in a keplerian disk our expression is equivalent to 
\begin{equation}
\nu_t = \beta' r^2 \Omega \qquad \hbox{with}\quad \beta' = {3 \over 2} 
\beta .
\end{equation} 
Such a prescription has been suggested originally by
Lynden-Bell and Pringle (1974), and recently it was used again by Duschl et al. (1998).
As a test, it is being applied to the modelling of accretion discs in active galactic nuclei
(Hur\'e \&  Richard 1999).

The reader may wonder why we have only used experimental results dating from the thirties,
namely those of Wendt (1933) and Taylor (1936).
The reason is that no one, since them, has studied in such extent the regime of outward
increasing angular momentum.\footnote{For instance, the Reynolds numbers explored by Coles (1965) are
one order of magnitude lower than those reached by Wendt, which explains why he did not encounter
the finite amplitude instability of neutral flows.}
We suspect that it is because the flow becomes then turbulent at once, 
 without undergoing a series of bifurcations associated with enticing
patterns. But we hope that experimentalists will turn again to this classical problem, 
which is 
of such great interest for geophysical and astrophysical fluid dynamics, and that they will 
explore the
rotation regimes for which the data are so incomplete.
In the meanwhile, the quest will continue to detect the finite amplitude instability
in computer simulations.

\begin{acknowledgements}
We thank our anonymous referee for his sharp remarks, which incited us to strengthen our case and to
dig even deeper into Wendt's experimental data.
\end{acknowledgements}

\bigskip

\noindent
{\bf References}

\begin{description}

\item Andereck C.D., Liu S.S., Swinney H.L. 1986, J. Fluid Mech. 164, 155

\item Balbus S.A., Hawley J.F 1991, ApJ 376, 214

\item  Balbus S.A., Hawley J.F., Stone J.M. 1996, ApJ 467, 76 

\item  Balbus S.A., Hawley J.F., Winters W.F. 1998, submitted to ApJ (astro-ph/9811057)

\item Chandrasekhar S. 1960, Proc. Nat. Acad. Sci. 46, 53

\item Coles D. 1965, J. Fluid Mech. 21, 385

\item Coughlin K., Marcus P. 1996, Phys. Rev. Letters. 77, 2214



\item Dubrulle B. 1993, Icarus 106, 59

\item Duschl W.J., Strittmatter P.A., Biermann P.L. 1998, AAS Meet. 192, \#66.17
 
\item Hur\'e J.-M., Richard D. 1999 (preprint)

\item Joseph D.D., Munson B.R. 1970, J. Fluid Mech. 43, 545

\item Kato S., Yoshizawa A., 1997, Publ. Astron. Soc. Japan, 49, 213

\item Lathrop D.P., Fineberg J., Swinney H.L. 1992, Phys. Rev. Letters., 68, 1515 

\item Lynden-Bell D., Pringle J.E. 1974, MNRAS 168, 603

\item Marcus P.S. 1984, J. Fluid Mech. 146, 45 \& 65
 
\item Rayleigh, Lord 1916, Proc. Roy. Soc. London A 93, 148

\item Serrin J. 1959, Arch. Ration. Mech. Anal. 3, 1

\item Shakura N.I., Sunyaev R.A. 1973, A\&A 24, 337

\item Taylor G.I. 1923, Phil. Trans. Roy. Soc. London A 223, 289

\item Taylor G.I. 1936, Proc. Roy. Soc. London A 157, 546

\item Wendt F., 1933, Ing. Arch.  4, 577

\item Zeldovich Y.B. 1981, Proc. Roy. Soc. London A 374, 299

\end{description}

\bigskip
\noindent
{\bf Appendix}
\bigskip

\noindent{\it On Zeldovich's analysis of Taylor's results.}
\medskip

Zeldovich (1981) had a similar goal in mind when he analyzed the results
of Taylor's experiment. But he started from the idea that the turbulent flow was
governed by the epicyclic frequency $N_\Omega$ and the turnover frequency $\omega$,
where
$$
N_\Omega^2 = {1 \over r^3} {d \over dr} (r^2 \Omega)^2
\quad \hbox{and} \quad \omega^2 = \left( r {d \Omega \over dr}\right)^2 ,
$$
since they measure respectively the stability of the flow and the strength of the shear.
He defines a non-dimensional parameter $Ty=N_\Omega^2/\omega^2$, 
akin to the Richardson number used in stratified shear flow,
and he seeks the confirmation of his intuition that $Ty$
be constant in Taylor's turbulent rotation profiles, in which case the angular
velocity would obey a power law
$$
\Omega \propto r^q.
$$
His best fit yields $q = 5.5$.
However these profiles are contaminated by the Ekman 
circulation mentioned above, and they differ marquedly from those
obtained by Wendt.

A consequence of this constant $Ty$ would be that the parameter $\alpha$ in (\ref{alpha})
would vary as 
$$
\alpha \propto \left(1 - {\Delta R \over R}\right)^{2q+4}
=  \left(1 - {\Delta R \over R}\right)^{15}
$$
a property which is not substantiated by the combined results of Taylor and Wendt,
as can be seen in Fig. 2.

\end{document}